\documentclass[reprint,a4paper]{revtex4-2}

\usepackage{graphicx}
\usepackage{xcolor}
\usepackage[colorlinks=true,linkcolor=blue,urlcolor=blue,citecolor=blue,bookmarks=true,bookmarksopenlevel=2]{hyperref}
\usepackage{amsmath}

\usepackage[utf8]{inputenc}
\usepackage[T1]{fontenc}
\usepackage{tgtermes}
\usepackage{microtype}
\usepackage{tabularx,booktabs}
\usepackage[referable]{threeparttablex}

\usepackage[normalem]{ulem}

\begin{document}

\title{De-excitation of $K$-shell hollow atoms with $12 \le Z \le 20$: transition rates and branching ratios}
\author{Karol Kozio{\l}}
\email{Karol.Koziol@ncbj.gov.pl}
\author{Jacek Rzadkiewicz} 
\email{Jacek.Rzadkiewicz@ncbj.gov.pl}
\affiliation{Narodowe Centrum Bada\'{n} J\k{a}drowych (NCBJ), Andrzeja So{\l}tana 7, 05-400 Otwock-\'{S}wierk, Poland}

\begin{abstract}
Investigating $K$-shell hollow atom spectra enhances our understanding of femtosecond phenomena in atomic physics, chemistry, and biology. 
Synchrotron measurements of two-electron one-photon (TEOP) transitions in low-$Z$ atoms have revealed discrepancies between experimental results and theoretical predictions of TEOP relative intensities. 
These discrepancies appear to originate from an incomplete description of an atom's response to the strong perturbation caused by $K$-shell double photoionization (DPI).
The multiconfiguration Dirac--Hartree--Fock relativistic configuration interaction method has been applied for studying the TEOP spectra of Mg, Al, Si, S, Ar, and Ca atoms.
The results show that branching ratios can be accurately reproduced by accounting for the effects of core and valence electron correlations, as well as the outer-shell ionization and excitation processes following $K$-shell DPI. 
\end{abstract}

\maketitle

\section{Introduction}

Atoms in which the outer shells are occupied while the innermost shell remains completely vacant are called $K$-shell hollow atoms. They provide a compelling environment for exploring the nature of exotic atomic states and the mechanisms responsible for their formation. 
These atoms can be generated through various physical processes, including nuclear decays and ion--atom collisions. 
Additionally, $K$-shell hollow atoms can arise from $K$-shell absorption of a single photon, followed by a purely quantum mechanical shake-off or a (quasi)classical knockout process \cite{Kanter2006,Huotari2008,Hoszowska2011}. 
Another pathway involves sequential multi-photon absorption occurring on a timescale comparable to the atom's decay time \cite{Young2010,Frasinski2013}. 
The latter process requires ultrashort, intense x-ray pulses, which can be generated using free electron lasers (FELs) \cite{slac-techreport,desy-techreport}. 

\begin{figure}
\includegraphics[width=\columnwidth]{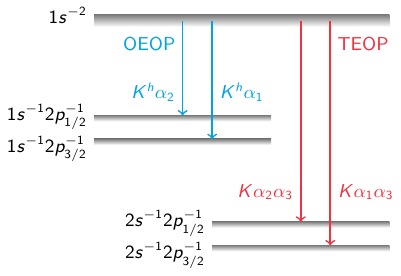}
\caption{\label{fig:oeop-teop}Level scheme (not to scale) showing the decay of $K$-shell hollow atoms via OEOP (blue arrows) and TEOP transitions (red arrows).}
\end{figure}

$K$-shell hollow atoms decay by non-radiative Auger or radiative transitions. 
The radiative transitions can occur via one-electron one-photon (OEOP) or via much less probable two-electron one-photon (TEOP) transitions (see Fig.~\ref{fig:oeop-teop}). 
In the case of the OEOP process, an electron jumps from the $2p$ sub-shell to the empty $K$ shell ($1s^{-2} \to 1s^{-1}2p^{-1}$), which is accompanied by a single x-ray photon emission. The resulting X-ray line is commonly labeled as $K^h\alpha_{1,2}$ or simply $K^h\alpha$. 
For the TEOP process, the empty $K$-shell is completely filled by simultaneous jumps of two electrons from the $L$ shell, one from the $2p$ sub-shell and one from the $2s$ sub-shell ($1s^{-2} \to 2s^{-1}2p^{-1}$), also with the emission of a single x-ray photon. The resulting X-ray line is commonly labeled as $K\alpha_{1,2}\alpha_3$ or simply $K\alpha\alpha^h$ or $K\alpha\alpha$. 
Both $K^h\alpha$ OEOP and $K\alpha\alpha$ TEOP transitions are sensitive to the relativistic, Breit interaction, and quantum electrodynamics (QED) effects \cite{Diamant2000a,Rzadkiewicz1999,Natarajan2008}. Because TEOP transitions are not allowed by the selection rules involving the one-configurational approach, their transition rates are very sensitive to the electron correlation effects \cite{Kadrekar2010,Kozio2017}. 
Moreover, the natural widths of the corresponding $K^h\alpha$ and $K\alpha\alpha$ lines give direct information on the $K$-shell hollow atoms lifetimes, which are one of the shortest lifetimes of any known bound atomic state \cite{Rzadkiewicz2005,Polasik2011a,Hoszowska2011,Hoszowska2013}.
Both $K\alpha\alpha^h$ TEOP and $K^h\alpha$ OEOP transitions originate from de-excitation of this same high-excited atomic state, so the branching ratio of the $K^h\alpha$ to $K\alpha\alpha^h$ transitions is a sensitive tool to test quantum-mechanics calculation approaches, because measured branching ratio does not depend on experimental circumstances such as $K$-shell ionization cross-section. 
Thus, TEOP and OEOP transitions provide a means to investigate both the fundamental principles of atomic physics and the characteristics of $K$-shell double photoionization (DPI) processes. 

$K^h\alpha$ OEOP transitions were observed for the first time by Charpak \cite{Charpak1953} and enlightened by Briand \textit{et al.} \cite{Briand1971}. 
Then they were observed further in many experiments~\cite{Hoszowska2010,Hoszowska2009a,Hoszowska2013,Keski-Rahkonen1977a,Mikkola1983,Soni1997,Verma2000,Oura2002,Raju2007,Ahopelto1979,Diamant2009,Diamant2009a,Diamant2000a,Diamant2003,Salem1984,Salem1982,Vukovic2001,Salem1983,Rzadkiewicz1999,Boschung1995,Salem1980}. 
The TEOP transitions were predicted for the first time by Heisenberg \cite{Heisenberg1925} in 1925, but only in 1975 the first experimental observation of TEOP transitions was reported by W{\"o}lfli \textit{et al.} \cite{Wolfli1975}. 
Till now, the TEOP transitions in $K$-shell hollow atoms have been investigated in many experiments~\cite{Knudson1976,Stoller1977,Schuch1979,Volpp1979,Tawara2002,Salem1982,Al-Ghazi1982,Isozumi1980}. They are also a subject of many theoretical considerations~\cite{Aberg1976a,Baptista1984,Saha2009,Gavrila1978,Safronova1977,Costa2006,Kadrekar2010,Wu2023}.

\begin{table*}[!htb]
\begin{threeparttable}
\caption{\label{tab:kaarates-mg}Total transition rates of $K\alpha_2\alpha_3$ transitions for Mg for various active spaces. The theoretical uncertainties are in parentheses.}
\begin{tabular*}{\linewidth}{@{\extracolsep{\fill}} l l cc ccc}
\toprule
\multicolumn{2}{l}{Active space for} & \multicolumn{2}{c}{No. of CSFs\tnotex{tab:kaarates-mg:a}} & \multicolumn{2}{c}{$K\alpha_2\alpha_3$ rate ($10^{10} \text{s}^{-1}$)} & \\
\cmidrule{3-4}\cmidrule{5-6}
\multicolumn{2}{l}{initial/final states} & $1s^{-2}$ & $2s^{-1}2p^{-1}$ & $A^L$ & $A^V$ & $A^L/A^V$ \\\midrule
AS0 & $2s^2 2p^6 3s^2$ / $1s^2 2s^1 2p^5 3s^2$ & 1 & 2 & 5.109 & 1.429 & 3.575 \\
AS1 & $\{ns\}$ / $\{ns,np\}$ ($n$=1-3) & 6 & 200 & 2.168 & 1.992 & 1.088 \\
AS2 & $\{ns,np\}$ ($n$=1-4)\tnotex{tab:kaarates-mg:b}  & 101 & 954 & 2.078 & 1.865 & 1.114 \\
AS3 & $\{ns,np\}$ ($n$=1-5)\tnotex{tab:kaarates-mg:b} & 233 & 2278 & 2.083 & 1.862 & 1.119 \\
AS4 & $\{ns,np\}$ ($n$=1-6)\tnotex{tab:kaarates-mg:b} & 420 & 4172 & 2.086(3) & 1.861(1) & 1.121 \\
\midrule
\multicolumn{2}{l}{Ref.~\cite{Kadrekar2010} (limited CI)} &&& 3.943$\times10^{-3}$ & 2.574 & 0.0015 \\
\multicolumn{2}{l}{Ref.~\cite{Kadrekar2010} (large CI)} &&& 2.262 & 2.169 & 1.043 \\
\multicolumn{2}{l}{Ref.~\cite{Saha2009}} &&& 5.11 &  &  \\
\midrule
\multicolumn{7}{l}{\textit{3s satellite}}\\
AS0 & $2s^2 2p^6 3s^1$ / $1s^2 2s^1 2p^5 3s^1$ & 1 & 6 & 5.871 & 1.641 & 3.578 \\
AS1 & $\{ns\}$ / $\{ns,np\}$ ($n$=1-3) & 8 & 493 & 2.258 & 2.228 & 1.014 \\
AS2 & $\{ns,np\}$ ($n$=1-4)\tnotex{tab:kaarates-mg:b}  & 213 & 2184 & 2.404 & 2.220 & 1.083 \\
AS3 & $\{ns,np\}$ ($n$=1-5)\tnotex{tab:kaarates-mg:b} & 500 & 5094 & 2.459 & 2.216 & 1.109 \\
AS4 & $\{ns,np\}$ ($n$=1-6)\tnotex{tab:kaarates-mg:b} & 909 &  9223 & 2.485(30) & 2.213(3) & 1.123 \\
\bottomrule
\end{tabular*}
\begin{tablenotes}
\item[a] \label{tab:kaarates-mg:a} States involving in $K\alpha\alpha$ and $K^h\alpha$ transitions only
\item[b] \label{tab:kaarates-mg:b} Excluding $1s$-$2p$ substitutions for initial states because of convergence issue
\end{tablenotes}
\end{threeparttable}
\end{table*}

%
%
%Recently the TEOP transitions following single-photon $K$-shell DPI have been observed in a highly accurate synchrotron experiment for Mg, Al and Si by Hoszowska \textit{et al.} \cite{Hoszowska2011}.
%In the experiment, the TEOP transition energies, branching ratios of the OEOP to TEOP transitions, and the TEOP linewidths were precisely measured.
%So far, this single-photon impact data provides the most reliable experimental results, which can rigorously test the most advanced atomic modeling.
%
%A comparison of the experimental values and theoretical predictions has shown a good agreement only for the TEOP energies.
%The experimental branching ratios of the OEOP to TEOP transitions and the TEOP linewidths are rather poorly reproduced by theory.
%The calculations based on second-order perturbation theory using single-electron screened hydrogenic wavefunctions \cite{Baptista1984} as well as the Hartree–Fock calculations based on a so-called `shake-down' model \cite{Aberg1976a} and multiconfiguration Dirac--Hartree--Fock (MCDHF) \cite{Saha2009} underestimate the experimental branching ratios \cite{Hoszowska2011} by a factor of 2--3.
%It is only the employment of the relativistic configuration interaction (RCI) formalism that allows reducing the discrepancies to the level of 15\%--30\% \cite{Kadrekar2010}. 
%Therefore, it is clear that accurate calculations of the TEOP intensities for low-$Z$ elements is called for \cite{Hoszowska2010}.

In our previous paper \cite{Kozio2017} we showed that quantitative reproduction of experimental branching ratios and $K\alpha\alpha^h$ linewidths for Mg, Al, and Si \cite{Hoszowska2011} is possible only on theoretical level including multiconfiguration Dirac--Hartree--Fock (MCDHF) plus relativistic configuration interaction (RCI) calculations and so-called Outer-shell Ionization and Excitation (OIE) effect. 
In present work we extend our study for next elements (S, Ar, and Ca) and show values for transition rates of $K\alpha\alpha^h$ and $K^h\alpha$ lines, including leading outer-shell satellites of these lines. Then we studied OIE effect influence on the branching ratio.

\section{Theoretical background}

\subsection{MCDHF-RCI method}

The calculations of radiative transition energies and rates have been carried out by means of the \textsc{Grasp2k} v1.1 \cite{Jonsson2013} code, based on the MCDHF method. 
The methodology of MCDHF calculations performed in the present study is similar to that published earlier, in many papers (see, e.g., \cite{Dyall1989,Grant2007}). 
The effective Hamiltonian for an $N$-electron system is expressed by
\begin{equation}
H = \sum_{i=1}^{N} h_{D}(i) + \sum_{j>i=1}^{N} C_{ij},
\end{equation}
where $h_D(i)$ is the Dirac operator for the $i$th electron and the terms $C_{ij}$ account for electron-electron interactions. In general, the latter is a sum of the Coulomb interaction operator and the transverse Breit operator. An atomic state function (ASF) with the total angular momentum $J$ and parity $p$ is assumed in the form
\begin{equation}
\Psi_{s} (J^{p} ) = \sum_{m} c_{m} (s) \Phi ( \gamma_{m} J^{p} ),
\end{equation}
where $\Phi ( \gamma_{m} J^{p} )$ are the configuration state functions (CSFs), $c_{m} (s)$ are the configuration mixing coefficients for state $s$, and $\gamma_{m}$ represents all information required to define a certain CSF uniquely. The CSFs are linear combinations of $N$-electron Slater determinants which are antisymmetrized products of 4-component Dirac orbital spinors. 
In present calculations, the initial and final states of considered transitions have been optimized separately and the biorthonormal transformation has been used for performing transition rates calculations \cite{Jonsson2007}. Following this, the so-called relaxation effect is taken into account. 
In the \textsc{Grasp2k} code, the Breit interaction contribution to the energy is added in perturbation way, after radial part of wavefunction is optimized. 
We calculated the Breit term in low-frequency limit (see, e.g., \cite{Kozio2018} for details), because frequency-dependent term is not appropriate for virtual orbitals \cite{Si2018}. 
Dependence of radiative transition rates on frequency-dependent Breit term is estimated to be below 0.01\% and it is negligible. 
Also two types of quantum electrodynamics (QED) corrections, self-energy (as screened hydrogenic approximation \cite{McKenzie1980} of data of Mohr and co-workers \cite{Mohr1992a}) and vacuum polarization (as potential of Fullerton and Rinker \cite{Fullerton1976}), have been included. 
The radiative transition rates were calculated in both velocity (Coulomb)~\cite{Grant1974} and length (Babushkin)~\cite{Babushkin1964} gauges. 

An accuracy of the wavefunction depends on the CSFs included in its expansion \cite{FroeseFischer2016}. The accuracy can be improved by extending the CSF set by including the CSFs originated by substitutions from orbitals occupied in the reference CSFs to unfilled orbitals of the active orbital set (Active Space, AS). 
The RCI method makes it possible to include the major part of the electron correlation contribution to the energy of the atomic levels and transition strengths. 
The difference between transition rates calculated in length and velocity gauges is a common way to test the quality of wavefunction obtained in self-consistent filed (SCF) calculations. 

\begin{figure}
\centering
\includegraphics[width=0.93\linewidth]{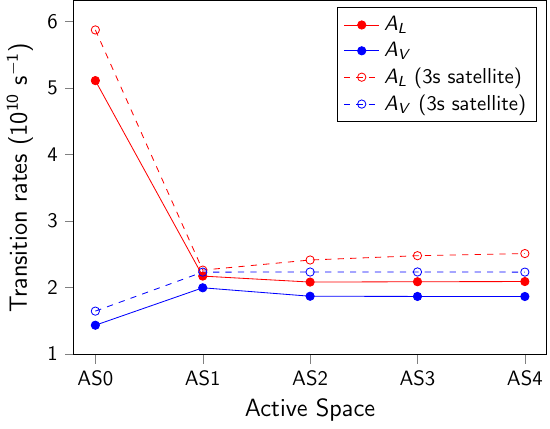}
\caption{\label{fig:mg-conv-rates}Convergence of calculations of $K\alpha_2\alpha_3$ transition rates for Mg.}
\end{figure}

Kadrekar and Natarajan found \cite{Kadrekar2010} that the discrepancies between $K\alpha\alpha^h$ transition rates calculated in length and velocity gauges may be reduced by using the RCI approach. They also found that the transition rates are very sensitive on the choice of the orbital set.
Our studies performed on a large number of test cases showed that using substitutions for ``hole'' orbitals in RCI expansion is crucial for producing reliable results that take into account important correlation effects. 
Hence, for initial states of $K\alpha\alpha^h$ transitions ($1s^{-2}$) the $\{ns\}$ ($n=1-3$) active space was used (containing CSFs produced by $1s$-$2s$ and $1s$-$3s$ substitutions) and for final states ($2s^{-1}2p^{-1}$) the $\{ns,np\}$ ($n=1-3$) active space was used (containing CSFs produced by substitutions for ``hole'' $2s$ and $2p$ orbitals). The RCI expansion has been created by using single (S) and double (D) substitutions from multireference set. In this way the length-velocity transition rate ratio, $I_{len}/I_{vel}$, may be reduced from 3.58-3.64 to 1.02-1.09. 
On the next stages we extend CSF set to $\{ns,np\}$ ($n=1-n_{max}$, $n_{max} = 4, 5$) active spaces, excluding only $1s$-$2p$ substitutions for initial states because of convergence issue. For $3p$ satellite of $K\alpha\alpha$ line of Ca we exclude also $n=3$ shell from substitutions for final states because of convergence issue. 

The theoretical uncertainties of radiative transition rates related to convergence with the size of a basis set have been estimated as rounded absolute value of difference between transition rates calculated within the two highest ASn for given model, i.e. $\delta A \approx |A(ASn)-A(ASn{-}1)|$ (assuming that correlation effects on transition rates are saturated).

\begin{figure}
\centering
\includegraphics[width=0.98\linewidth]{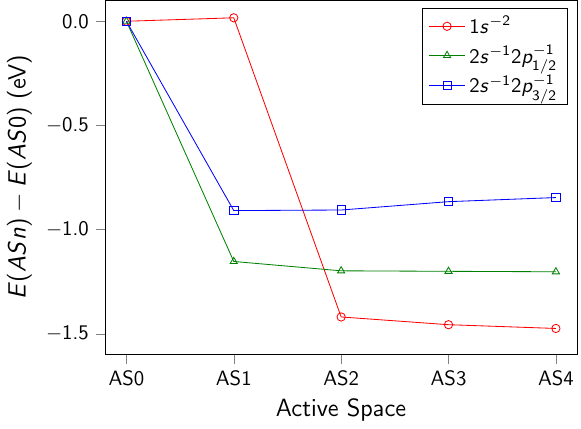}
\caption{\label{fig:mg-conv-energy}Convergence of calculations of $1s^{-2}$ and $2s^{-1}2p^{-1}$ states energy for Mg.}
\end{figure}

\subsection{Total shake probabilities}

The total shake probabilities, i.e., shake-off and shake-up, have been calculated by applying the sudden approximation model \cite{Carlson1968} and using MCDHF wave functions for two valence ionization scenarios, namely OIE1 and OIE2. 
The OIE1 corresponds to the ionization/excitation of the valence ($3s$ in the case of Mg, $3s$ and $3p$ in the cases of Al, Si, S, and Ar, and $3p$ and $4s$ in the case of Ca) electrons due to the sudden atomic potential change resulting from the single $K$-shell vacancy. 
In the case of the OIE2, the more pronounced potential change is caused by two $K$-shell vacancies due to the quasi-simultaneous removal of two $1s$ electrons (for details, see \cite{Polasik2011a}).

\section{Results}

\subsection{$K\alpha\alpha^h$ transition rates for Mg -- controlling SCF convergence}

\begin{table}[!htb]
\caption{\label{tab:kaaenergy-mg}Energy of $1s^{-2}$, $2s^{-1}2p_{1/2}^{-1}$, $2s^{-1}2p_{1/2}^{-1}$ states and $K\alpha_1\alpha_3$ and $K\alpha_2\alpha_3$ transitions for Mg for various active spaces.}
\tabcolsep=0.5\tabcolsep
\begin{tabular*}{\linewidth}{@{\extracolsep{\fill}} l ccccc}
\toprule
Active & \multicolumn{5}{c}{Energy (eV)}\\
\cmidrule{2-6}
space & $1s^{-2}$ & $2s^{-1}2p_{3/2}^{-1}$ & $2s^{-1}2p_{1/2}^{-1}$ & $K\alpha_1\alpha_3$ & $K\alpha_2\alpha_3$\\\midrule
AS0 & -2663.80 & -5264.66 & -5249.28 & 2600.86 & 2585.46\\
AS1 & -2663.79 & -5265.57 & -5250.43 & 2601.78 & 2586.64\\
AS2 & -2665.22 & -5265.57 & -5250.48 & 2600.34 & 2585.25\\
AS3 & -2665.26 & -5265.53 & -5250.48 & 2600.27 & 2585.22\\
AS4 & -2665.28 & -5265.51 & -5250.48 & 2600.23 & 2585.20\\
\bottomrule
\end{tabular*}
\end{table}

\begin{table}[!htb]
\begin{threeparttable}
\caption{\label{tab:kaarates-al}Total transition rates of $K\alpha\alpha=K\alpha_1\alpha_3+K\alpha_2\alpha_3$ transitions (statistically averaged per initial $1s^{-2}$ state) for Al for various active spaces. The theoretical uncertainties are in parentheses.}
\tabcolsep=0.5\tabcolsep
\begin{tabular*}{\linewidth}{@{\extracolsep{\fill}} l cc ccc}
\toprule
Active & \multicolumn{2}{c}{No. of CSFs} & \multicolumn{2}{c}{$K\alpha\alpha$ rate ($10^{10} \text{s}^{-1}$)} & \\
\cmidrule{2-3}\cmidrule{4-5}
space & $1s^{-2}$ & $2s^{-1}2p^{-1}$ & $A^L$ & $A^V$ & $A^L/A^V$ \\\midrule
AS0 & 2 & 17 & 6.736 & 1.865 & 3.611 \\
AS1 & 24 & 806 & 2.643 & 2.598 & 1.017 \\
AS2 & 939 & 7707 & 2.314 & 2.297 & 1.008 \\
AS3 & 2776 & 22381 & 2.535(230) & 2.382(90) & 1.064 \\
\midrule
\multicolumn{3}{l}{Ref.~\cite{Kadrekar2010} (limited CI)} & 2.441$\times10^{-3}$ & 2.169 & 0.011 \\
\multicolumn{3}{l}{Ref.~\cite{Kadrekar2010} (large CI)} & 3.248 & 2.854 & 1.138 \\
\multicolumn{3}{l}{Ref.~\cite{Kadrekar2010} (large CI\tnotex{tab:kaarates-al:a})} & 2.929 & 2.949 & 0.993 \\
\midrule
\multicolumn{6}{l}{\textit{3s satellite}}\\
AS0 & 4 & 35 & 7.239 & 2.007 & 3.606 \\
AS1 & 37 & 1406 & 2.898 & 2.822 & 1.027 \\
AS2 & 1806 & 11688 & 2.984 & 2.801 & 1.065 \\
AS3 & 5174 & 32826 & 3.033(50) & 2.689(120) & 1.128 \\
\multicolumn{6}{l}{\textit{3p satellite}}\\
AS0 & 1 & 2 & 7.707 & 2.134 & 3.612 \\
AS1 & 6 & 200 & 3.016 & 2.565 & 1.176 \\
AS2 & 101 & 954 & 2.950 & 2.430 & 1.214 \\
AS3 & 229 & 2278 & 3.029(80) & 2.463(40) & 1.230 \\
\bottomrule
\end{tabular*}
\begin{tablenotes}
\item[a] \label{tab:kaarates-al:a} Not including coupling with the 3p electron
\end{tablenotes}
\end{threeparttable}
\end{table}

\begin{table}[!htb]
\caption{\label{tab:kaarates-si}Total transition rates of $K\alpha\alpha=K\alpha_1\alpha_3+K\alpha_2\alpha_3$ transitions (statistically averaged per initial $1s^{-2}$ state) for Si for various active spaces. The theoretical uncertainties are in parentheses.}
\tabcolsep=0.5\tabcolsep
\begin{tabular*}{\linewidth}{@{\extracolsep{\fill}} l cc ccc}
\toprule
Active & \multicolumn{2}{c}{No. of CSFs} & \multicolumn{2}{c}{$K\alpha\alpha$ rate ($10^{10} \text{s}^{-1}$)} & \\
\cmidrule{2-3}\cmidrule{4-5}
space & $1s^{-2}$ & $2s^{-1}2p^{-1}$ & $A^L$ & $A^V$ & $A^L/A^V$ \\\midrule
AS0 & 5 & 40 & 8.452 & 2.323 & 3.638 \\
AS1 & 57 & 956 & 3.487 & 3.332 & 1.047 \\
AS2 & 2262 & 16889 & 2.989(500) & 2.780(560) & 1.075 \\
\midrule
\multicolumn{3}{l}{Ref.~\cite{Kadrekar2010} (limited CI)} & 4.759$\times10^{-2}$ & 4.092 & 0.012  \\
\multicolumn{3}{l}{Ref.~\cite{Kadrekar2010} (large CI)} & 3.489 & 3.378 & 1.033 \\
\multicolumn{3}{l}{Ref.~\cite{Saha2009}} & 8.37 &  &  \\
\midrule
\multicolumn{6}{l}{\textit{3s satellite}}\\
AS0 & 8 & 76 & 9.052 & 2.492 & 3.632 \\
AS1 & 75 & 1706 & 3.990 & 3.517 & 1.134 \\
AS2 & 2812 & 24179 & 3.840(150) & 3.199(320) & 1.201 \\
\multicolumn{6}{l}{\textit{3p satellite}}\\
AS0 & 2 & 17 & 8.546 & 2.348 & 3.639 \\
AS1 & 24 & 806 & 3.460 & 3.085 & 1.121 \\
AS2 & 939 & 7707 & 3.503(50) & 3.047(40) & 1.150 \\
\bottomrule
\end{tabular*}
\end{table}

\begin{table}[!htb]
\caption{\label{tab:kaarates-s}Total transition rates of $K\alpha\alpha=K\alpha_1\alpha_3+K\alpha_2\alpha_3$ transitions (statistically averaged per initial $1s^{-2}$ state) for S for various active spaces. The theoretical uncertainties are in parentheses.}
\tabcolsep=0.5\tabcolsep
\begin{tabular*}{\linewidth}{@{\extracolsep{\fill}} l cc ccc}
\toprule
Active & \multicolumn{2}{c}{No. of CSFs} & \multicolumn{2}{c}{$K\alpha\alpha$ rate ($10^{10} \text{s}^{-1}$)} & \\
\cmidrule{2-3}\cmidrule{4-5}
space & $1s^{-2}$ & $2s^{-1}2p^{-1}$ & $A^L$ & $A^V$ & $A^L/A^V$ \\\midrule
AS0 & 5 & 40 & 12.579 & 3.425 & 3.673 \\
AS1 & 57 & 340 & 4.900 & 4.995 & 0.981 \\
AS2 & 2295 & 17313 & 4.273(630) & 4.311(690) & 0.991 \\
\midrule
\multicolumn{3}{l}{Ref.~\cite{Saha2009}} & 12.4 &  &  \\
\midrule
\multicolumn{6}{l}{\textit{3s satellite}}\\
AS0 & 8 & 76 & 13.397 & 3.653 & 3.668 \\
AS1 & 75 & 834 & 6.017 & 5.116 & 1.176 \\
AS2 & 3960 & 26733 & 5.764(260) & 4.840(280) & 1.191 \\
\multicolumn{6}{l}{\textit{3p satellite}}\\
AS0 & 5 & 51 & 12.459 & 3.389 & 3.676 \\
AS1 & 69 & 693 & 5.532 & 5.047 & 1.096 \\
AS2 & 3066 & 22033 & 5.338(200) & 4.693(360) & 1.137 \\
\bottomrule
\end{tabular*}
\end{table}

\begin{table}[!htb]
\caption{\label{tab:kaarates-ar1}Total transition rates of $K\alpha_2\alpha_3$ transitions for Ar for various active spaces. The theoretical uncertainties are in parentheses.}
\tabcolsep=0.5\tabcolsep
\begin{tabular*}{\linewidth}{@{\extracolsep{\fill}} l cc ccc}
\toprule
Active & \multicolumn{2}{c}{No. of CSFs} & \multicolumn{2}{c}{$K\alpha_2\alpha_3$ rate ($10^{10} \text{s}^{-1}$)} & \\
\cmidrule{2-3}\cmidrule{4-5}
space & $1s^{-2}$ & $2s^{-1}2p^{-1}$ & $A^L$ & $A^V$ & $A^L/A^V$ \\\midrule
AS0 & 1 & 2 & 16.580 & 4.478 & 3.702 \\
AS1 & 6 & 12 & 6.050 & 6.926 & 0.873 \\
AS2 & 105 & 1026 & 6.222(180) & 6.858(70) & 0.907 \\
\midrule
\multicolumn{3}{l}{Ref.~\cite{Kadrekar2010} (limited CI)} & 3.116 & 8.193 & 0.380 \\
\multicolumn{3}{l}{Ref.~\cite{Kadrekar2010} (large CI)} & 5.271 & 5.707 & 0.924 \\
\multicolumn{3}{l}{Ref.~\cite{Saha2009}} & 16.5 &  &  \\
\midrule
\multicolumn{6}{l}{\textit{3s satellite}}\\
AS0 & 1 & 6 & 17.313 & 4.673 & 3.705 \\
AS1 & 8 & 78 & 7.136 & 6.498 & 1.098 \\
AS2 & 253 & 3102 & 7.914(780) & 6.610(120) & 1.197 \\
\multicolumn{6}{l}{\textit{3p satellite}}\\
AS0 & 2 & 17 & 18.290 & 4.982 & 3.672 \\
AS1 & 24 & 106 & 6.884 & 7.054 & 0.976 \\
AS2 & 968 & 8115 & 6.339(550) & 6.232(830) & 1.017 \\
\bottomrule
\end{tabular*}

\caption{\label{tab:kaarates-ar2}Total transition rates of $K\alpha_1\alpha_3$ transitions for Ar for various active spaces. The theoretical uncertainties are in parentheses.}
\tabcolsep=0.5\tabcolsep
\begin{tabular*}{\linewidth}{@{\extracolsep{\fill}} l cc ccc}
\toprule
Active & \multicolumn{2}{c}{No. of CSFs} & \multicolumn{2}{c}{$K\alpha_1\alpha_3$ rate ($10^{7} \text{s}^{-1}$)} & \\
\cmidrule{2-3}\cmidrule{4-5}
space & $1s^{-2}$ & $2s^{-1}2p^{-1}$ & $A^L$ & $A^V$ & $A^L/A^V$ \\\midrule
AS0 & 1 & 2 & 36.077 & 5.369 & 6.720 \\
AS1 & 6 & 12 & 8.508 & 12.973 & 0.656 \\
AS2 & 105 & 1026 & 8.868(360) & 12.148(830) & 0.730 \\
\midrule
\multicolumn{3}{l}{Ref.~\cite{Kadrekar2010} (limited CI)} & 0.787 & 11.04 & 0.071 \\
\multicolumn{3}{l}{Ref.~\cite{Kadrekar2010} (large CI)} & 7.959 & 7.983 & 0.997 \\
\multicolumn{3}{l}{Ref.~\cite{Saha2009}} & 35.7 &  &  \\
\bottomrule
\end{tabular*}
\end{table}

\begin{table}[!htb]
\begin{threeparttable}
\caption{\label{tab:kaarates-ca1}Total transition rates of $K\alpha_2\alpha_3$ transitions for Ca for various active spaces. The theoretical uncertainties are in parentheses.}
\tabcolsep=0.5\tabcolsep
\begin{tabular*}{\linewidth}{@{\extracolsep{\fill}} l cc ccc}
\toprule
Active & \multicolumn{2}{c}{No. of CSFs} & \multicolumn{2}{c}{$K\alpha_2\alpha_3$ rate ($10^{10} \text{s}^{-1}$)} & \\
\cmidrule{2-3}\cmidrule{4-5}
space & $1s^{-2}$ & $2s^{-1}2p^{-1}$ & $A^L$ & $A^V$ & $A^L/A^V$ \\\midrule
AS0 & 1 & 2 & 25.606 & 6.895 & 3.714 \\
AS1 & 10 & 754 & 8.279 & 8.590 & 0.964 \\
AS2 & 331 & 3762 & 7.609(670) & 7.859(740) & 0.968 \\
\midrule
\multicolumn{3}{l}{Ref.~\cite{Kadrekar2010} (limited CI)} & 0.5881 & 10.84 & 0.054 \\
\multicolumn{3}{l}{Ref.~\cite{Kadrekar2010} (large CI)} & 6.460 & 7.081 & 0.912 \\
\multicolumn{3}{l}{Ref.~\cite{Saha2009}} & 22.0 &  &  \\
\midrule
\multicolumn{6}{l}{\textit{3p satellite}}\\
AS0 & 2 & 17 & 24.492 & 6.679 & 3.667 \\
AS1\tnotex{tab:kaarates-ca1:a} & 44 & 2450 & 10.234 & 10.323 & 0.991 \\
AS2\tnotex{tab:kaarates-ca1:a}  & 3562 & 11998 & 9.160(1080) & 9.201(1130) & 0.996 \\
\multicolumn{6}{l}{\textit{4s satellite}}\\
AS0 & 1 & 6 & 23.489 & 6.308 & 3.724 \\
AS1 & 18 & 2533 & 9.094 & 9.193 & 0.989 \\
AS2 & 887 & 11440 & 8.704(390) & 8.803(390) & 0.989 \\
\bottomrule
\end{tabular*}
\begin{tablenotes}
\item[a] \label{tab:kaarates-ca1:a} Excluding also $n=3$ shell from substitutions for final states because of convergence issue
\end{tablenotes}
\end{threeparttable}

\caption{\label{tab:kaarates-ca2}Total transition rates of $K\alpha_1\alpha_3$ transitions for Ca for various active spaces. The theoretical uncertainties are in parentheses.}
\tabcolsep=0.5\tabcolsep
\begin{tabular*}{\linewidth}{@{\extracolsep{\fill}} l cc ccc}
\toprule
Active & \multicolumn{2}{c}{No. of CSFs} & \multicolumn{2}{c}{$K\alpha_1\alpha_3$ rate ($10^{8} \text{s}^{-1}$)} & \\
\cmidrule{2-3}\cmidrule{4-5}
space & $1s^{-2}$ & $2s^{-1}2p^{-1}$ & $A^L$ & $A^V$ & $A^L/A^V$ \\\midrule
AS0 & 1 & 2 & 12.048 & 1.982 & 6.080 \\
AS1 & 10 & 754 & 0.855 & 1.238 & 0.690 \\
AS2 & 331 & 3762 & 0.744(120) & 1.060(180) & 0.702 \\
\midrule
\multicolumn{3}{l}{Ref.~\cite{Kadrekar2010} (limited CI)} & 0.2931 & 3.429 & 0.085 \\
\multicolumn{3}{l}{Ref.~\cite{Kadrekar2010} (large CI)} & 2.201 & 2.286 & 0.963 \\
\multicolumn{3}{l}{Ref.~\cite{Saha2009}} & 10.2 &  &  \\
\bottomrule
\end{tabular*}
\end{table}

In Table~\ref{tab:kaarates-mg} we show the transition rates of the $K\alpha_2\alpha_3$ transitions for Mg. 
The theoretical uncertainties of radiative transition rates are shown for the final (largest) AS. 
The $K\alpha_1\alpha_3$ transition has a rate $\sim10^4$ times smaller than the $K\alpha_2\alpha_3$ transition and the $K^h\alpha_1$ transition has a rate $\sim10^3$ times smaller than the $K^h\alpha_2$ transition, thus their contributions to BR are negligible.  
As one can see in the table, the $K\alpha\alpha^h$ transition rates in length ($A^L$) and velocity ($A^V$) forms significantly differ when the virtual orbital contributions are neglected (AS0).
Next, using only S substitutions to virtual orbitals does not improve the ratio $I_{len}/I_{vel}$.
In order to get convergence and an agreement between $A^L$ and $A^V$, one has to extend the calculations to the SD substitutions with an active set at least up to $n=4$ (AS2).
The lower part of Table~\ref{tab:kaarates-mg} presents the predictions for the $3s$ satellites of the ``pure'' TEOP that, due to the outer-shell ionization and excitation (OIE \cite{Polasik2011a}) processes, can noticeably modify the effective (observed in an experiment) transition rates and the corresponding TEOP linewidths, as in the case of OEOP transitions~\cite{Polasik2011a}.

As one can see from Fig.~\ref{fig:mg-conv-rates}, for both ``pure'' and $3s$ satellite $K\alpha\alpha^h$ lines the calculations for transition rates converge well if at least AS2 stage is used. Similar convergence is obtained for $1s^{-2}$ and $2s^{-1}2p^{-1}$ states energy (see Fig.~\ref{fig:mg-conv-energy}) and, as consequence, for $K\alpha_1\alpha_3$ and $K\alpha_2\alpha_3$ lines energy (see Table~\ref{tab:kaaenergy-mg}). 
It is worth to noticing that the $K\alpha\alpha^h$ lines energy, in contrast to $K\alpha\alpha^h$ transition rates, is only weekly dependent on the level of electron correlation taken into account. The difference between energy of $K\alpha_2\alpha_3$ transition calculated on AS0 and AS4 stages is 0.26~eV and the difference between energy of $K\alpha_2\alpha_3$ transition calculated on AS2 and AS4 stages is 0.05~eV. In the case of weaker $K\alpha_1\alpha_3$ transition the similar numbers are 0.63~eV and 0.11~eV, respectively. 

\begin{table*}[!htb]
\caption{\label{tab:kharates}Total transition rates of $K^h\alpha_{1,2}$ transitions for selected elements.}
\begin{tabular*}{\linewidth}{@{\extracolsep{\fill}} l l cc cccc}
\toprule
& & \multicolumn{2}{c}{$K^h\alpha$ rate ($10^{13}\ \text{s}^{-1}$)} & \multicolumn{4}{c}{Other theoretical ($10^{13}\ \text{s}^{-1}$)}\\
\cmidrule{3-4}\cmidrule{5-8}
Atom & Transition & $A^L$ & $A^V$ & Ref.~\cite{Natarajan2008} & Ref.~\cite{Saha2009} & Ref.~\cite{Costa2006} & Ref.~\cite{Chen1991}\\\midrule
Mg & $K^h\alpha_{1,2}$ & 4.758 & 4.438 & 5.47 & 4.74 & &\\
& \ldots 3s satellite & 4.764 & 4.444 &&&&\\
Al & $K^h\alpha_{1,2}$ & 6.785 & 6.373 &  & 6.75 & 6.92&\\
& \ldots 3s satellite & 6.791 & 6.379 &&&&\\
& \ldots 3p satellite & 6.813 & 6.400 &&&&\\
Si & $K^h\alpha_{1,2}$ & 9.395 & 8.883 & 10.32 & 9.65 & &\\
& \ldots 3s satellite & 9.454 & 8.939 &&&&\\
& \ldots 3p satellite & 9.421 & 8.907 &&&&\\
S & $K^h\alpha_{1,2}$ & 16.778 & 16.006 &  & 17.0 & &\\
& \ldots 3s satellite & 16.869 & 16.092 &&&&\\
& \ldots 3p satellite & 16.826 & 16.051 &&&&\\
Ar & $K^h\alpha_{2}$ & 27.525 & 26.422 & 29.29 & 26.4 & & 30.7\\
& $K^h\alpha_{1}$ & 0.364 & 0.350 & 0.399 & 0.590 && 0.430\\
& $K^h\alpha_{1,2}$\ 3s satellite & 27.895 & 26.777 &&&&\\
& $K^h\alpha_{1,2}$\ 3p satellite & 27.889 & 26.771 &&&&\\
Ca & $K^h\alpha_{2}$ & 42.555 & 41.099 & 44.87 & 39.7 & &\\
& $K^h\alpha_{1}$ & 1.186 & 1.145 & 1.29 & 1.22 &&\\
& $K^h\alpha_{1,2}$\ 3p satellite & 43.747 & 42.251 &&&&\\
& $K^h\alpha_{1,2}$\ 4s satellite & 43.696 & 42.202 &&&&\\
\bottomrule
\end{tabular*}
\end{table*}

\subsection{$K\alpha\alpha^h$ and $K^h\alpha$ transition rates for Al, Si, S, Ar, and Ca}

Similar calculations as for Mg have been performed for OEOP and TEOP transitions for Al, Si, S, Ar, and Ca atoms. The numbers are presented in Tables~\ref{tab:kaarates-al}, \ref{tab:kaarates-si}, \ref{tab:kaarates-s}, \ref{tab:kaarates-ar1}, \ref{tab:kaarates-ar2}, \ref{tab:kaarates-ca1}, and \ref{tab:kaarates-ca2}. 
To ensure that our calculations take a reasonable time and obtain a reasonable accuracy, we kept the calculations to the AS3 stage for Al and the AS2 stage for Si, S, Ar, and Ca atoms. 
Because of mixing the CSFs involving $2s^{-1}2p^{-1}$ hole states within ASFs, the $K\alpha_1\alpha_3$ and $K\alpha_2\alpha_3$ transition rates are considered together in all case where there are more than two $2s^{-1}2p^{-1}$ final states of $K\alpha\alpha^h$ transitions. 
Since there are a lot of atomic levels originating from a given spectator hole for open-shell atomic systems, the statistically averaged rates per initial $1s^{-2}$ state are presented. 
Table~\ref{tab:kharates} collects total transition rates of $K^h\alpha_{1,2}$ transitions. It has been found that $K^h\alpha_{1,2}$ transitions are less sensitive to CI calculations. 
The $A^L$ numbers are quoted from \cite{Kozio2019a}. 

A good agreement between $A^L$ and $A^V$ indicates the quality of the ASF representations \cite{Froese-Fischer1997}. 
In considered case, the $A^L/A^V$ ratio is in the range 0.907--1.121 in the case of ``pure'' $K\alpha_2\alpha_3$ line. 
For satellite transitions, for which the convergence is more difficult to achieve, the $A^L/A^V$ ratio is in the range 0.989--1.230. 
In the case of hypersatellite transitions the $A^L/A^V$ ratio is in the ranges 1.045--1.077 and 1.038--1.067 for ``pure'' $K^h\alpha_2$ line and its satellite transitions, respectively.

One can also see that the transition rates for the $3s$, $3p$, and $4s$ satellites of the $K\alpha\alpha^h$ transitions are about 20\%–35\% higher than those for the diagram ones. It was also found that in the case of the $3s$, $3p$, and $4s$ satellites of OEOP ($K^h\alpha$) transitions, the change in the corresponding transition rates is significantly lower (below 1\%). Thus, it is clear that $3s$, $3p$, and $4s$ OIE processes can modify the branching ratios.

\subsection{$K^h\alpha$ to $K\alpha\alpha^h$ branching ratio}

The $K^h\alpha$ and $K\alpha\alpha^h$ BR has been calculated by using the following expression
\begin{equation}
BR = \frac{I(K^h\alpha)}{I(K\alpha\alpha^h)} = \frac{\sum_{ij} A_{ij}}{\sum_{ik} A_{ik}}\;,
\label{eq:br-simple}
\end{equation}
where $A_{ij}$ and $A_{ik}$ are rates for transition between $i$th $1s^{-2}$ initial states and $j$th $1s^{-1}2p^{-1}$ ($K^h\alpha$ transitions) or $k$th $2s^{-1}2p^{-1}$ final states ($K\alpha\alpha^h$  transitions), respectively. 
The OIE (shake) processes that change electronic configuration of de-excited $K$-shell hollow atoms, modify their radiative transition rates. In order to take into account this affection on the BR we have used the following equation
\begin{equation}
BR = \frac{I_0(K^h\alpha) + \sum_s I_s(K^h\alpha) \frac{I_s}{I_0}}{I_0(K\alpha\alpha^h) + \sum_s I_s(K\alpha\alpha^h) \frac{I_s}{I_0}} \;,
\label{eq:br-oie}
\end{equation}
where the intensity ratio $I_s/I_0$ of the intensity of the main $K^h\alpha$ or $K\alpha\alpha^h$ line (i.e. without additional spectator hole), $I_0$, to the intensity of the its $nl$-shell satellite, $I_s^{nl}$, is given according to binomial distribution \cite{McGuire1973}:
\begin{equation}
\frac{I_s^{nl}}{I_0} = \frac{N^{nl}P_{ion}^{nl}}{1-P_{ion}^{nl}} \;,
\label{eq:binom}
\end{equation}
where $P_{ion}^{nl}$ is an ionization probability for the $nl$ shell (due to shake processes) and $N^{nl}$ is a number of electrons occurring on the $nl$ shell.

The calculated values of total shake probabilities (in percent per subshell) for the OIE1 and OIE2 scenarios presented in Table~\ref{tab:shake} have been used for $I_s/I_0$ factor calculations in the Eq.~\eqref{eq:binom}.

\begin{table}[!htb]
\caption{\label{tab:shake}Total shake probabilities (in \% per subshell) as a~result of single (OIE1) and double (OIE2) $K$-shell ionization. Results for Mg, Al, and Si have been presented previously in Ref.~\cite{Kozio2017}.}
\begin{tabular*}{\linewidth}{@{\extracolsep{\fill}} llrr}
\toprule
Atom & Subshell & OIE1 & OIE2\\
\midrule
Mg &  $3s$  & 20.62 & 49.21\\
Al &  $3s$  & 11.81 & 33.65\\
 &  $3p$  & 15.08 & 37.09\\
Si &  $3s$  & 7.89 & 24.71\\
 &  $3p$  & 18.24 & 44.62\\
Si &  $3s$  & 4.39 & 15.13\\
 &  $3p$  & 19.68 & 57.76\\
Ar &  $3s$  & 2.85 & 10.3\\
 &  $3p$  & 18.47 & 55.89\\
Ca &  $3p$  & 10.73 & 36.12\\
 &  $4s$ & 20.34 & 49.09\\
\bottomrule
\end{tabular*}
\end{table}

\begin{table*}[!htb]
\begin{threeparttable}
\caption{\label{tab:br-rat}The $K^h\alpha$ to $K\alpha\alpha^h$ branching ratios for Mg, Al, Si, S, Ar, and Ca calculated by using various approaches. The theoretical uncertainties are in parentheses.}
\begin{tabular*}{\linewidth}{@{\extracolsep{\fill}} ll ccc ccc}
\toprule
& gauge & Mg & Al & Si & S & Ar & Ca\\\midrule
MCDHF 	& length & 909\tnotex{tab:br-rat:a} & 1002\tnotex{tab:br-rat:a} & 1105\tnotex{tab:br-rat:a}  & 1329 & 1626 & 1648\\
		& velocity & 3102\tnotex{tab:br-rat:a} & 3409\tnotex{tab:br-rat:a} & 3807\tnotex{tab:br-rat:a}  & 4668 & 5798 & 5938\\
RCI 	& length & 2273(4) & 2644(240)\tnotex{tab:br-rat:a} & 3096(518)\tnotex{tab:br-rat:a}  & 3848(567) & 4369(126) & 5619(494)\\
		& velocity & 2365(2) & 2646(100)\tnotex{tab:br-rat:a} & 3148(634)\tnotex{tab:br-rat:a}  & 3639(582) & 3790(39) & 5202(490)\\
RCI+OIE1  	&  length  & 2202(12) & 2562(109) & 3010(240) & 3729(270) & 4435(155) & 5536(302)\\
   			&  velocity  & 2304(2) & 2633(56) & 3121(303) & 3646(303) & 3958(168) & 5205(300)\\
RCI+OIE2  	&  length  & 2122(11) & 2442(99) & 2872(219) & 3503(239) & 4375(151) & 5344(282)\\
   			&  velocity  & 2220(1) & 2586(54) & 3041(288) & 3564(290) & 4040(175) & 5053(283)\\
Recommended  &   & 2122(99) & 2442(175) & 2872(277) & 3503(247) & 4375(367) & 5344(405) \\
\midrule
Experiment: &&&&&&&\\
Ref.~\cite{Hoszowska2011} && 1838(258) & 2115(403) & 2610(370) &  &  & \\
Theory: &&&&&&&\\
Ref.~\cite{Aberg1976a} && 667 & 758 & 833&&&\\
Ref.~\cite{Gavrila1978} &&  & 682 &  &  &  & 1240 \\
Ref.~\cite{Safronova1977} &&  &  &  &  &  & 1770 \\
Ref.~\cite{Baptista1984} && 576.5 & 742.0 & 906.0 &  &  &  \\
Ref.~\cite{Costa2006} &&  & 686 & &&&\\
Ref.~\cite{Saha2009} && 928 & 999 & 1126 & 1370 & 1632 & 1851 \\
(Ref.~\cite{Kadrekar2010} + Ref.~\cite{Natarajan2008})\tnotex{tab:br-rat:b}\tnotex{tab:br-rat:c} && 2417 & 2359\tnotex{tab:br-rat:d}/2617\tnotex{tab:br-rat:e} & 3007 &  & 5624 & 7121 \\
\bottomrule
\end{tabular*}
\begin{tablenotes}
\item[a] \label{tab:br-rat:a} Reported previously in Ref.~\cite{Kozio2017}
\item[b] \label{tab:br-rat:b} Values presented in Ref.~\cite{Hoszowska2011} for Mg, Al, Si
\item[c] \label{tab:br-rat:c} $K^h\alpha$ rates are from Ref.~\cite{Natarajan2008} and $K\alpha\alpha^h$ rates are interpolated from Ref.~\cite{Kadrekar2010}, both in the length gauge
\item[d] \label{tab:br-rat:d} $K\alpha\alpha^h$ rates include the coupling of the initial and final state vacancies with the $3p$ electron
\item[e] \label{tab:br-rat:e} $K\alpha\alpha^h$ rates include without the coupling of the initial and final state vacancies with the $3p$ electron
\end{tablenotes}
\end{threeparttable}
\end{table*}

\begin{figure}
	\includegraphics[width=\columnwidth]{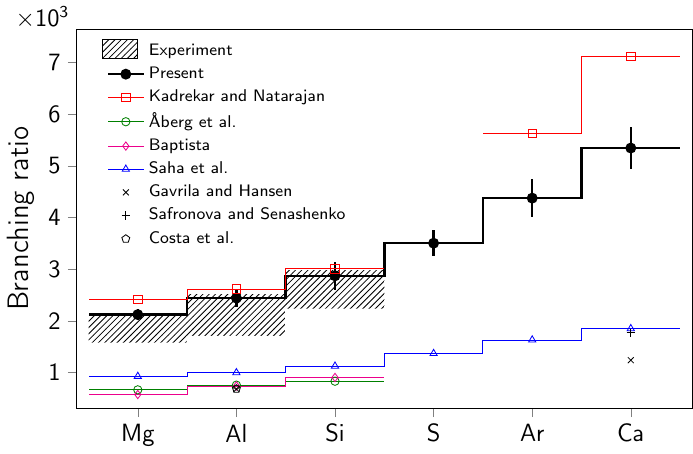}
	\caption{\label{fig:br-rat}The $K^h\alpha$ to $K\alpha\alpha^h$ branching ratios for Mg, Al, Si, S, Ar, and Ca calculated by using various approaches and compared to available experimental data. Sources: experiment -- Ref.~\cite{Hoszowska2011}; Kadrekar and Natarajan -- Ref.~\cite{Kadrekar2010} + Ref.~\cite{Natarajan2008}; {\AA}berg et al. -- Ref.~\cite{Aberg1976a}; Baptista -- Ref.~\cite{Baptista1984}; Saha et al. -- Ref.~\cite{Saha2009}; Gavrila and Hansen -- Ref.~\cite{Gavrila1978}; Safronova and Senashenko -- Ref.~\cite{Safronova1977}; Costa et al. -- Ref.~\cite{Costa2006}.}
\end{figure}

In Table~\ref{tab:br-rat} the branching ratios for Mg, Al, Si, S, Ar, and Ca, calculated using various approaches, including OIE1 and OIE2, are presented and compared to the experimental values and previous theoretical predictions. The graphical representation of these data is shown on Fig.~\ref{fig:br-rat}. 
The theoretical uncertainties are originated from the uncertainties of $K\alpha\alpha^h$ transition rates. 
One can see that employing the RCI and OIE1 approach improves distinctly the branching ratios over the simple MCDHF model. The branching ratio values calculated by using the RCI+OIE2 model with the length gauge reproduces the measured branching ratios better than any other theoretical predictions published so far. 
One can see that the inclusion of the $3s$ and $3p$ OIE2 contribution to the branching ratios reduces the discrepancies between the experiment and theory for Mg, Al, and Si. 
The improvement is achieved at a relatively low cost related to the increase of differences between the length and velocity gauge branching ratio calculations within the RCI+OIE2 model by up to 6\%. This is a natural consequence of taking into account the satellite TEOP transitions between states having in general more open subshells than those for the ``diagram'' ones. Nevertheless, only this approach can provide an atomic model that is able to take into account the OIE processes that can strongly affect TEOP transitions. 

The recommended BR values in Table~\ref{tab:br-rat} are based on the RCI+OIE2 numbers calculated by using the length gauge. The theoretical uncertainties for these values are a combination of the uncertainties for RCI+OIE2(length) values and the difference between RCI+OIE2(length) and RCI+OIE2(velocity) numbers. 

\section{Conclusions}

In conclusion, we have shown that employing the MCDHF-RCI calculations with OIE corrections enables to reproduce the experimental branching ratios for Mg, Al, and Si. 
The obtained theoretical values are in agreement within 10\%--14\% of the measured branching ratios for Mg, Al, and Si. 
For S, Ar, and Ca we provide theoretical predictions for branching ratios, that may be examined by future experiments. 
The results of our studies set new theoretical limits for the TEOP transitions in the low-$Z$ atomic range. 
We also hope that this work will guide future theoretical studies for higher-$Z$ elements and new experiments with a higher accuracy.

%\bibliographystyle{model1-num-names}
%\bibliography{references,refs2}

\end{document}